\documentclass[journal=jacsat,manuscript=article]{achemso}

\usepackage[version=3]{mhchem} 
\usepackage{times}
\usepackage{bm}
\usepackage{amsmath,amssymb}

\author{Sathwik Bharadwaj}
\affiliation{Elmore Family School of Electrical and Computer Engineering, Purdue University, West Lafayette, Indiana 47907, USA.}
\altaffiliation{Purdue Quantum Science and Engineering Institute, Purdue University, West Lafayette, Indiana 47907, USA.}
\author{Zubin Jacob}
\affiliation{Elmore Family School of Electrical and Computer Engineering, Purdue University, West Lafayette, Indiana 47907, USA.}
\altaffiliation{Purdue Quantum Science and Engineering Institute, Purdue University, West Lafayette, Indiana 47907, USA.}
\email{zjacob@purdue.edu}
\title[An \textsf{achemso} demo]{Unraveling Optical Polarization at Deep Microscopic Scales in Crystalline Materials}

\abbreviations{Optical Polarization, Lattice Nonlocal Optical Response, Crystalline materials, Polarization Texture}

\begin{document} 







\begin{abstract}
Nanophotonics, the study of light-matter interaction at scales smaller than the wavelength of radiation, has widespread applications in plasmonic waveguiding, topological photonic crystals, super-lensing, solar absorbers, and infrared imaging. The physical phenomena governing these effects can be described using a macroscopic homogenized refractive index. However, the lattice-level description of optical polarization in a crystalline material using a quantum theory has been unresolved. Inspired by the dynamics of electron waves and their corresponding band structure, we propose a microscopic optical band theory of solids specifically applicable to optical polarization. This framework reveals propagating waves hidden deep within a crystal lattice. These hidden waves arise from crystal-optical-indices, a family of quantum functions obeying crystal symmetries, and cannot be described by the conventional concept of refractive index.  We present for the first time - the hidden waves and deep microscopic optical band structure of 14 distinct materials. We choose Si, Ge, InAs, GaAs, CdTe, and others from Group IV, III-V, and II-VI due to their technological relevance but our framework can be extended to a wide range of emerging 2D and 3D materials. In contrast to the macroscopic refractive index of these materials used widely today, this framework shows that hidden waves exist throughout the crystal lattice and have unique optical polarization texture and crowding. We also present an open-source software package, Purdue-Picomax, for the research community to discover hidden waves in new materials like hBN, graphene, and Moire materials. Our work establishes a foundational crystallographic feature to discover novel deep microscopic optical waves in light-matter interaction.

\end{abstract}


The conventional concepts of optical polarization and refractive index are foundational to quantifying the light-matter interaction of materials from a macroscopic point of view \cite{horsley2021zero, Defo_Molesky_Rodriguez_2022, kim2016highly, Darrick_maximum_refractive_index, shim2021fundamental}. This frequency-dependent function, $n(\omega)$ takes the common Drude-Lorentz or hydrodynamic form and is used widely in multiple fields from microphotonics \cite{reserbat2021quantum}, metamaterials \cite{orlov2011engineered}, nanophotonics \cite{khurgin2017landau} to quantum optics \cite{yan2015projected}. For example, optical fiber modes are described by the refractive index of glass \cite{petersen2014mid} whereas silicon waveguides are modeled using the refractive index of silicon \cite{almeida2004all}. On the other hand, optical waves in topological photonic crystals and hyperbolic media (eg: CaCo$_3$,hBN) are governed by their spatially periodic or anisotropic refractive index, respectively \cite{jacob2014hyperbolic}. Even in quantum optics, the homogenized refractive index is used to understand the enhancement or suppression of spontaneous emission of emitters placed inside a medium \cite{Halas}.  It is therefore pertinent to ask what role crystal symmetries play in light-matter interactions and whether a description shift emerges beyond this widely studied nanoscale limit. 

We emphasize that deep microscopic local field effects do not just change the quantitative values for the refractive index but have been proven to encode topological physics related to nonlocal optical phases of matter.  These microscopic effects also capture hidden topological properties of crystalline materials such as the optical $N$-invariant \cite{ToddSathwik_opticalN}. Such quantum numbers are connected to microscopic optical polarizability and are distinct from topological invariants known from electronic band theory. This forms an important motivation for studying microscopic polarization texture since it can encode optical skyrmions at deep sub-wavelength scales. The winding number and twists in deep microscopic quantum polarization are unique phenomena that are distinct from skyrmions in plasmonic lattices or photonic crystals as well as free space light beams. Similarly, the quantum gyro-electric effect and effective photon mass in Maxwell Hamiltonians \cite{sathwik_picophotonics} also require microscopic polarization functions as input. Thus, a complete quantum theory is necessary to understand the optical polarization at deep microscopic scales.

In a quantum theory, the optical polarization has two distinct flavors: a longitudinal response related to the Coulombic field, and a transverse response tied to the propagating electromagnetic waves. The electronic properties and band structure are affected primarily by the longitudinal Coulombic field. Thus the bulk of existing research in density functional theory has focused on the microscopic understanding of the longitudinal density-density correlation function \cite{Adler, NWiser, varas2016quantum}. In stark contrast, the above-mentioned topological optical polarization effects, deep sub-wavelength optical skyrmions, and Maxwell Hamiltonians necessarily require the transverse optical response function of matter. We note that the conventional density-density correlation does not capture the optical response of a material to a transverse electromagnetic field, especially in the high momentum regime including microscopic local field effects. This issue was resolved only recently by introducing the current-current correlation function which describes transverse propagating waves in the deep microscopic regime \cite{sathwik_picophotonics}. 

The early attempts to quantify the refractive index using microscopic fields were taken into account using the Clausius–Mossotti relation (Lorentz–Lorenz equation) \cite{Hannay_1983, Rysselberghe}, which substitutes a homogeneous cavity for the simple cubic lattice of polarizable atomic sites. This results in a relationship between the molecular polarizability and the macroscopic refractive index \cite{Aspnes}. However, the Clausius-Mossotti relation does not account for frequency or momentum dependency in the refractive index, nor does it take into account the symmetry of the Brillouin zone of materials. It should be noted that the widely accepted approximation of using polarizable harmonic oscillators instead of atoms is restricted to the classical regime. Recent attention has also fallen on nonlocal optical effects that arise from modifications to the refractive index in ultra-subwavelength systems \cite{bondarev2018optical, yang2019general, baumberg2019extreme}, where the nonlocal modification is considered in the continuum limit of hydrodynamic light-matter interaction \cite{teperik2013quantum, garcia2008nonlocal, gonccalves2020plasmon, fitzgerald2016quantum}. Multiple interesting works have studied quantum and nonlocal effects of plasmons using Fermi liquid theory \cite{lundeberg2017tuning, hofmann2015plasmon}. Distinct from the continuum limit, our work here focuses on crystalline symmetries that naturally occur at the atomistic level. We focus on transparent non-metallic insulating materials and study the deep microscopic eigenwaves with large momenta that occur within the material bandgap. 

The current work also has connections to the field of hyperbolic media where large momentum bulk eigenwaves lead to unique sub-diffraction effects \cite{zheng2022molding, ni2021long, galfsky2015active, ma2021ghost}. It is natural to ask - what is the fundamental limit to extreme momentum optical waves inside a crystalline material? The conventional low momentum optical waves and the light line only exist near the high symmetry $\Gamma$ point of a crystalline medium. Our quantum response function approach shows that bulk eigenwaves can adopt the symmetry of the lattice leading to extreme momentum optical waves that are both non-planar and inhomogeneous. We focus on energy scales within the bandgap of a transparent insulating material to overcome detrimental effects such as Landau damping of plasmons and hyperbolic polaritons.  Our results are also relevant to atomic-scale light confinements in sub-nanometre size cavities which have recently gained significant interest \cite{Babar2023-nt}.


In the static limit, the contemporary understanding of macroscopic polarization was developed by King-Smith and Vanderbilt \cite{Vanderbilt_King-Smith, Resta_review}, which asserts that the polarization difference between two crystal states is determined by the geometric quantum phase of the electronic wave function, and the macroscopic polarization encodes the global topology through the Berry connection \cite{watanabe2018inequivalent}. We note that the macroscopic polarization framework describes a system that lacks inversion symmetry when subjected to static fields and is not intended to treat the optical polarization at the lattice level. It has been a long-standing challenge to understand the dynamics of optical waves at deep microscopic scales in crystalline materials probed beyond static or uniform field response \cite{SipeMahon, talebi2018electron, wunsch2006dynamical, markel2016introduction, ToddSathwik_opticalN}. We show that the deep microscopic optical band structure described here fully addresses this challenge. We demonstrate that the momentum exchange processes that occur at sub-nanometer (nm) length scales in a crystal determine the deep microscopic lattice optical polarization. 

\begin{figure}
    \centering
    \includegraphics[width = 5.5in]{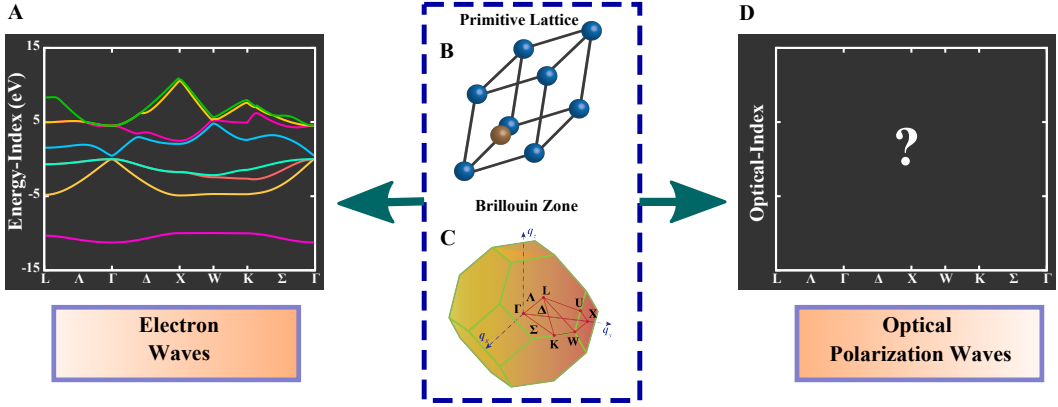}
    \caption{{\bf Deep microscopic light-matter interaction in crystalline materials}.  The concept of refractive index captures the optical properties of materials. However, it does not capture the crystal symmetries and hidden polarization texture deep inside the periodic lattice. Inspired by the concept of electronic band theory, we unravel the deep microscopic optical band structure of a crystalline material and discover hidden polarization waves. The schematic shows a representative primitive lattice in a Zinc-Blend crystal, corresponding to the electronic band structure (B) and Brillouin zone (C).}
    \label{fig:figure1}
\end{figure}

Our solution is inspired by the band structure of electron waves in a crystal, which is central to materials physics, and provides a foundational classification of materials based on the energy-momentum-wavefunction $(E-\bm{k}-\psi)$ relationship. The electronic band structure represents the allowed energy indices for the electron waves within the Brillouin zone of a material (Figure~\ref{fig:figure1}(A)). A similar band theory has also been developed for phonons which has led to considerable advances in nanoscale heat transport and thermal materials discovery \cite{qian2021phonon}.  In contrast, here we put forth the deep microscopic optical polarization band theory of crystalline materials which reveals hidden waves in a crystal lattice throughout the Brillouin zone (Figure~\ref{fig:figure1}(D)). 



We rigorously define optical polarization and bulk transverse eigenwaves using a quantum generalization of the refractive index. Specifically, in this article, we show the existence of two distinct hidden forms of polarization waves in a crystal: a) deep microscopic optical transverse waves, and b) deep microscopic optical longitudinal waves. These are captured by our optical band theory which describes light-matter interaction at the level of crystal lattice.  We unravel the deep microscopic optical band structure and hidden waves of 14 Zinc-Blend and diamond-type cubic materials. Here, we choose Si, Ge, InAs, GaAs, CdTe, and others from Group IV, III-V, and II-VI due to their technological relevance. We note that our framework can be easily extended to include hexagonal boron nitride, graphene, Moire materials, MoS$_2$, etc. 

Along with building theoretical foundations, we also put forth a software package, Purdue-Picomax, to unravel microscopic optical waves in a wide range of 2D and 3D materials. These deep microscopic optical bands serve as symmetry indicators beyond electronic or phononic properties and simultaneously provide a signature for identifying optical topological phases in natural materials. \mbox{Purdue-Picomax} automates our theoretical framework and enables calculations of unique physical properties of 2D and 3D materials unavailable in existing light-matter interaction theories: symmetry-indicating lattice conductivity (SILC), deep microscopic optical transverse and longitudinal polarization band structure. We expect Purdue-Picomax to aid the exploration of optical materials and optical topological phenomena with unique polarization texture at the lattice level \cite{sathwik_picophotonics}. 

\section{Deep Microscopic Optical Waves in Crystalline Materials}
In this section, we develop the theoretical framework to unravel hidden optical waves of materials starting from the corresponding basic lattice parameters. Our approach for obtaining optical polarization encodes the minimal coupling between the material and the gauge field $\bm{A}_\mu \equiv (\phi, \bm{A}_i)$, which satisfies the local U(1) gauge symmetry. Our crucial insight is that the gauge fields have to obey additional crystalline symmetries which leads to novel deep microscopic light-matter interaction. We note that the central physical quantity in determining the deep microscopic optical band structure of a material is the lattice conductivity tensor ${\bm{\sigma}}(\bm{r}, \bm{r}', \omega)$. Through a Fourier expansion in the reciprocal lattice space, we define the symmetry-indicating lattice conductivity (SILC) tensor of the form $\bm{\sigma}^{\bm{G}\,{\bm{G}}'}_{\omega\,\bm{q}}$, where $\omega$ is the frequency, $\bm{q}$ is the crystal-wavevector of the optical polarization waves, $\bm{G}, \bm{G'}$ are the reciprocal lattice vectors, and $\bm{q}$ is restricted within the Brillouin zone of the material. 

The lattice conductivity satisfies the invariance under space group operations ($\bm{\mathcal{G}}$), given by
\begin{equation}\label{eq:symmetryconductivity}
\bm{\sigma}^{\bm{G}\,{\bm{G}}'}_{\omega\,\bm{q}_t} = e^{i(\bm{G}'-\bm{G})\cdot\bm{\tau}_{\bm{\mathcal{G}}}}\,\bm{\sigma}^{\bm{G}_t\,{\bm{G}_t}'}_{\omega\,\bm{q}},
\end{equation}
where, $\bm{q}_t = \bm{\mathcal{G}}\cdot\bm{q} + \bm{G}_{\bm{\mathcal{G}}}$, $\bm{G}_t = \bm{\mathcal{G}}^{-1}\cdot(\bm{G}+\bm{G}_{\bm{\mathcal{G}}})$, and $\bm{G}_{\bm{\mathcal{G}}}$ is the auxiliary reciprocal lattice vector introduced to include the class of nonsymmorphic crystals. The symmetry groups of $\bm{q}$ are different along different symmetry axes of the Brillouin zone, and we see from Eq.~\ref{eq:symmetryconductivity} that the structure of the lattice conductivity tensor is dictated by the symmetry group of $\bm{q}$ (see Supplementary Material). 

The polarization fields in a crystal lattice form a completely orthogonal set and satisfy the Bloch form. This is in stark contrast to the conventional continuum theory used to describe the optical properties of solids \cite{ermolaev2021giant}. We put forth the definition of the optical polarization waves as the eigenfunctions of the SILC for a given frequency and momentum $(\omega, \bm{q})$:
\begin{equation}\label{eq:spectraldecomposition}
\bm{\sigma}^{\bm{G}\,{\bm{G}}'}_{\bm{q}\,\omega} = -i\omega\,\sum_{\lambda=0}^{D_f-1} \frac{\left({\alpha_{\lambda\,\omega\,\bm{q}}-1}\right)}{4\pi}\,\frac{\left[\bm{\Tilde{p}}^{\dagger}_{\lambda\,\omega\,\bm{q}}\,\otimes\,\bm{\Tilde{p}}_{\lambda\,\omega\,\bm{q}}\right]}{\left|\bm{\Tilde{p}}_{\lambda\,\omega\,\bm{q}}\right|^2},
\end{equation}
where, $\otimes$ represents the Cartesian outer product, $D_f$ is the number of deep microscopic optical bands considered, the eigenvector \mbox{$\bm{\Tilde{p}}_{\lambda\,\omega\,\bm{q}} = \left[\bm{u}_{\lambda\,\omega\,\bm{q}\,\bm{G}_0}, \bm{u}_{\lambda\,\omega\,\bm{q}\,\bm{G}_1}, ... \right]$}, and $\bm{u}_{\lambda\,\omega\,\bm{q}\,\bm{G}_i}$ represents the plane-wave components of the optical polarization wave. 

We define the optical indices of a crystal as the set of eigenvalues $\alpha_{\lambda\,\omega\,\bm{q}}$, which generalize the classical concept of the refractive index. For a given optical-index $\alpha_{\lambda\,\omega\,\bm{q}}$, we put forth the expression for the optical polarization wave, given by 
\begin{equation}
\bm{p}_{\lambda\,\omega\,\bm{q}}(\bm{r}, \omega) = \frac{\left(\alpha_{\lambda\,\omega\,\bm{q}}-1\right)}{4\pi} \,e^{i\bm{q}\cdot\bm{r}}\,\sum_{\bm{G}} e^{i\bm{G}\cdot\bm{r}}\, \bm{u}_{\lambda\,\omega\,\bm{q}\,\bm{G}}.
\end{equation}
Furthermore, we define the exponentially localized real-space deep microscopic optical functions ($\bm{W}_{\lambda\omega\bm{R}}(\bm{r})$) to decompose the polarization waves (up to a gauge transformation), 
\begin{equation}
\bm{p}_{\lambda\,\omega\,\bm{q}}(\bm{r}, \omega) = \frac{\left(\alpha_{\lambda\,\omega\,\bm{q}}-1\right)}{4\pi}\,\int_{\rm cell}\,d\bm{r}\,\sum_{\bm{R}}\bm{W}_{\lambda\omega\bm{R}}(\bm{r})\,e^{-i\bm{q}\cdot\left(\bm{r}-\bm{R}\right)}.
\end{equation}
where, $\bm{R}$ is lattice vector in the real space. We note that our real-space deep microscopic optical functions are analogous to Wannier functions known for the real-space representation of electronic Bloch functions. For a given frequency $\omega$, the set of $\alpha_{\lambda\,\omega\,\bm{q}}$ obtained within the Brillouin zone of the crystal describes the deep microscopic optical band structure with $\lambda$ being the band-index. The orthogonality of the deep microscopic optical functions at distinct crystal lattice sites offers a useful foundation for the expansion of optical polarization waves. We note that, unlike
the traditional electronic Wannier functions, the deep microscopic optical functions are vectorial and represent the fluctuating optical polarization waves. 

At the $\rm \Gamma$ point, we notice that the space group symmetry of the crystal is preserved. However, away from the $\rm \Gamma$ point, $\bm{q}$-symmetry group will be a subgroup of this space group. The compatibility relations between the irreducible representations of the symmetry groups at different points in the Brillouin zone determine the degeneracy pattern of the optical indices $\alpha_{\lambda\,\omega\,\bm{q}}$, and the optical indices serve as symmetry indicators. In principle, the number of optical bands $D_f$ is infinite, however, it can be truncated to a suitably large $\bm{G}$ vector. We emphasize that if we consider only a single optical band ($D_f = 1$), in the zero-momentum limit, the square root of the optical index ($\sqrt{\alpha_{\lambda = 0\,\omega\,\bm{q}\rightarrow 0}}$) reduces to the macroscopic refractive index $n(\omega)$. Thus we retrieve the conventional optical theory as a limiting case of our polarization framework. In general the optical indices ($\alpha_{\lambda\,\omega\,\bm{q}}$) are complex, and the imaginary part of optical indices represents the lattice extinction coefficient index. 

\subsection*{Deep Microscopic Optical-indices}
In this section, we unravel the deep microscopic optical band structures of crystalline materials, each of which represents a distinct set of optical indices. We note that an electromagnetic pulse carries a vector potential $\bm{A}$ perpendicular to $\bm{q}$, and the corresponding properties of the deep microscopic optical transverse band structure $\left\{\alpha^T_{\lambda\,\omega\,\bm{q}}, \bm{p}_{\lambda\,\omega\,\bm{q}}^T\right\}$ completely characterize the lattice optical response of a crystal from a light pulse. In contrast, the deep microscopic optical longitudinal band structure $\left\{\alpha^L_{\lambda\,\omega\,\bm{q}}, \bm{p}_{\lambda\,\omega\,\bm{q}}^L\right\}$ governs the electrical screening properties of crystalline materials. We define the transverse and longitudinal optical polarization waves to satisfy the following conditions:
\begin{align}
\bm{\nabla}\cdot\bm{p}_{\lambda\,\omega\,\bm{q}}^T = 0;\ \bm{\nabla}\times\bm{p}_{\lambda\,\omega\,\bm{q}}^L = 0.     
\end{align}

\begin{figure}
    \centering
    \includegraphics[width = 5.5in]{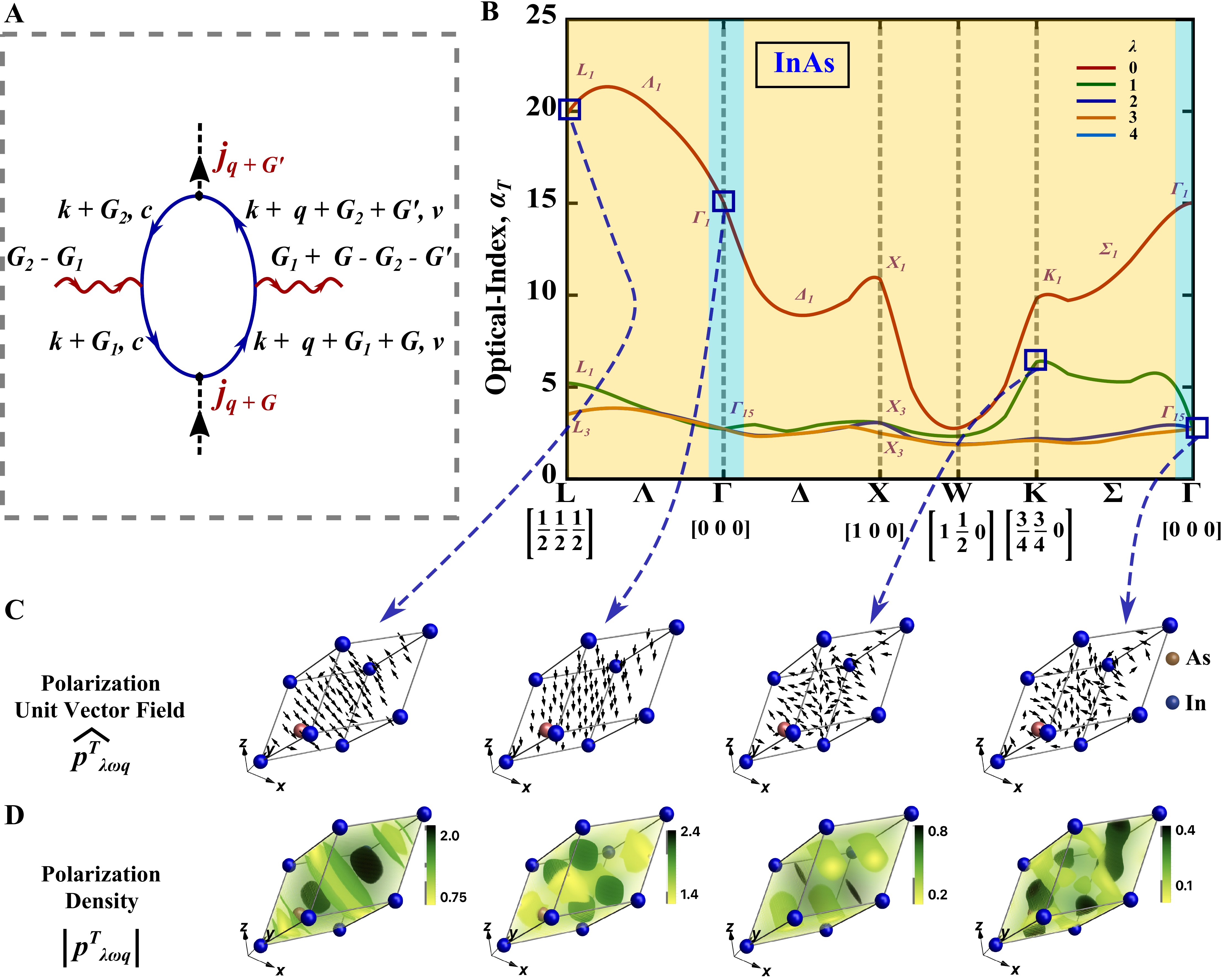}
    \caption{{\bf Unraveling hidden waves and deep microscopic optical waves (transverse) of a crystalline material}. We observe a significant texture and crowding of the hidden waves even within the primitive cell of a Zinc-Blend material. (A) The Feynman diagram depicts the symmetry indicating lattice transverse conductivity of a crystalline material responsible for inducing the optical transverse polarization waves. The dashed lines represent the incoming ($\bm{j}_{\bm{q}+\bm{G}}$) and outgoing ($\bm{j}_{\bm{q}+\bm{G}'}$) transverse current density, internal solid lines are the electron propagators of the conduction ($c$) and valence ($v$) bands, and wiggly lines are the momentum exchange processes with the lattice. Here, $\bm{G},\bm{G}',\bm{G}_1,\bm{G}_2$ are the discrete reciprocal lattice vectors, and $\bm{q}$, $\bm{k}$ are restricted to the first Brillouin zone. (B) deep microscopic optical transverse band structure of InAs at $10\,$THz. The labels on the deep microscopic optical bands correspond to the irreducible representations of the $\bm{q}$-symmetry group. The blue-shaded region around the $\Gamma$-point shown here is the domain accessed by classical optics. (C) Hidden optical transverse polarization unit vector field distribution at a few high-symmetry points plotted within the primitive cell of InAs. (D) Hidden optical transverse polarization density in units of $e^-/\text{\AA}^2$ corresponding to (C) are plotted within a primitive cell of InAs.}
    \label{fig:figure2}
\end{figure}

In Fig.~\ref{fig:figure2}(A) and Fig.~\ref{fig:figure3}(A), we have displayed the first-order Feynman diagrams corresponding to the symmetry-indicating lattice transverse and longitudinal conductivity, respectively. The wiggly lines in the Feynman diagram represent the momentum exchange processes with the lattice. The momentum exchange processes occur at sub-nm length scales ($1/|\bm{G}| \sim a$, where $a$ is the lattice constant), and these momentum exchange processes result in the deep microscopic optical band structure of a crystal lattice. When we neglect these momentum exchange processes, the dimension of this set reduces to unity and the optical-index is simply the dielectric function of the material. 

An electromagnetic pulse carries a transverse vector potential which induces the SILC component defined by the current-current correlation function. Following the Feynman Diagram in Fig.~\ref{fig:figure2}(A), we can derive the expression for $\left(\bm{\sigma}^T\right)^{\bm{G}\,{\bm{G}}'}_{\omega\,\bm{q}}$, given by   
\begin{align}\label{eq:transversecond}
\left(\bm{\sigma}^T\right)^{\bm{G}\,{\bm{G}}'}_{\omega\,\bm{q}} = &\left(\frac{-2i\,e^2}{\Omega\, \omega}\right)\times\nonumber\\&\sum_{c,v,\bm{k}}\Bigg[\frac{\left<{c,\bm{k}}\right|e^{-i\left(\bm{G}+\bm{q}\right)\cdot\bm{r}}\,\bm{t}_{\bm{G}}\cdot\bm{J}_0\left|{v,\bm{k}+\bm{q}}\right>\left<{v,\bm{k}+\bm{q}}\right|\,e^{i\left(\bm{G}'+\bm{q}\right)\cdot\bm{r}'}\,\bm{t}_{\bm{G}'}\cdot\bm{J}_0\left|{c,\bm{k}}\right>}{\left(\epsilon_{c\bm{k}}-\epsilon_{v\bm{k}+\bm{q}}+\hbar\omega+i\hbar\alpha\right)}+c.c\Bigg],  \end{align}
where, $\bm{J}_0$ is the current density operator, $\bm{t}_{\bm{G}}$ is the unit vector component perpendicular to $(\bm{q}+\bm{G})$, $\left|c,\bm{k}\right>$ and $\left|v,\bm{k}+\bm{q}\right>$ are the electronic Bloch functions of conduction and valence bands, $\Omega$ is the crystal volume, and $\epsilon_{c\bm{k}}$ and $\epsilon_{v\bm{k}+\bm{q}}$ are the electronic eigen-energy of conduction and valence bands, respectively. 

The longitudinal component of the SILC provides the density-density correlation induced by a scalar potential perturbation, and the corresponding expression can be derived using a similar approach, given by
\begin{align}\label{eq:longitudinalcond}
\left(\bm{\sigma}^L\right)^{\bm{G}\,{\bm{G}}'}_{\omega\,\bm{q}}=\Bigg(\frac{2i\,e^2\omega}{\Omega}&\,\frac{1}{|\bm{q}+\bm{G}||\bm{q}+\bm{G}'|}\Bigg)\times \nonumber\\ & \sum_{c,v,\bm{k}}\left[\frac{\left<c,\bm{k}\right|e^{-i\left(\bm{q}+\bm{G}\right)\cdot\bm{r}}\left|v,\bm{k}+\bm{q}\right>\left<v,\bm{k}+\bm{q}\right|e^{i\left(\bm{q}+\bm{G}'\right)\cdot\bm{r}'}\left|c,\bm{k}\right>}{\left(\epsilon_{c\bm{k}}-\epsilon_{v\bm{k}+\bm{q}}+\hbar\omega+i\hbar\alpha\right)} + c.c \right].   
\end{align}

{Lattice Nonlocal Ohm's Law}: A distinctive feature of the optical polarization waves is the emergence of the lattice nonlocal Ohm's law within a crystal lattice. Conventionally, dynamical Ohm's law in a crystalline solid is given by $\bm{J}(\omega) = \sigma(\omega) \bm{E}(\omega)$, where $\bm{J}$ is the cell-averaged continuum current density. However, the optical transverse polarization waves satisfy the lattice nonlocal Ohm's law in a crystalline lattice, given by
\begin{equation}
\bm{J}^T_{\lambda\,\omega\,\bm{q}} = -i\omega\,\frac{\left(\alpha^T_{\lambda\,\omega\,\bm{q}}-1\right)}{4\pi} \,e^{i\bm{q}\cdot\bm{r}}\,\sum_{\bm{G}} e^{i\bm{G}\cdot\bm{r}}\,\bm{u}^T_{\lambda\,\omega\,\bm{q}\,\bm{G}}. 
\end{equation}
In the continuum limit, our expression reduces to the classical form of Ohm's law. Hence, the optical polarization waves generalize the classical Ohm's law within a crystal lattice. 
\subsection*{Deep Microscopic Hidden Waves}
The central outcome of our analysis is the unraveling of hidden waves through our deep microscopic optical band structure of 14 distinct Zinc-Blend and diamond-type materials. In Fig.~\ref{fig:figure2}(B), we use our theory to solve for the deep microscopic optical transverse band structure in InAs at frequency $\omega_0 = 10\,$THz. It is immediately revealed that there exist hidden waves deep within the crystal lattice that cannot be captured by the classical wave theory. The blue-shaded region in Fig.~\ref{fig:figure2}(B), represents the low momentum domain of the deep microscopic optical band structure which recovers the classical optics. At the $\Gamma$-point, the square root of the first optical transverse index $\sqrt{\alpha^T_{0\,\omega_0\,\Gamma}}$ reduces to the classical refractive index, hence recovering the conventional optical theory. Furthermore, in a departure from conventional wisdom, the first optical transverse index $\alpha^T_{0\,\omega\,\bm{q}}$ has a maximum at a momentum along the $\Lambda$ axis (Fig.~\ref{fig:figure2}(B)).  We recognize this as a striking feature specific to the deep microscopic optical transverse band structure arising from the ultra sub-wavelength momentum exchange processes occurring within the lattice. 


In Fig.~\ref{fig:figure2}(C) and Fig.~\ref{fig:figure2}(D), we have plotted the unit vector field distribution and density of the hidden waves at $10\,$THz for a few selected bands and high-symmetry points. These plots display the hidden polarization texture and crowding within the unit cell of the material. This phenomenon is beyond the classical description of light-matter interaction where polarization only varies slowly in space.  We see that the first optical transverse polarization state around the $\Gamma$-point (considered along $\Delta$ direction), $\bm{p}_{0\,\omega\,\,\bm{q}}^T$ is aligned along the direction perpendicular to $\hat{\bm{q}} = [100]$, representing the classical limit. In Fig.~\ref{fig:figure2}(C) and Fig.~\ref{fig:figure2}(D) (also movie S1), we observe that away from the $\Gamma$-point, the hidden waves have a considerable polarization texture and crowding at the ultra-subwavelength level. The inhomogeneous distribution of density of the optical polarization waves is attributed to the pico-scopic momentum exchange processes with the lattice as described earlier.
\begin{figure}
    \centering
    \includegraphics[width = 5.5in]{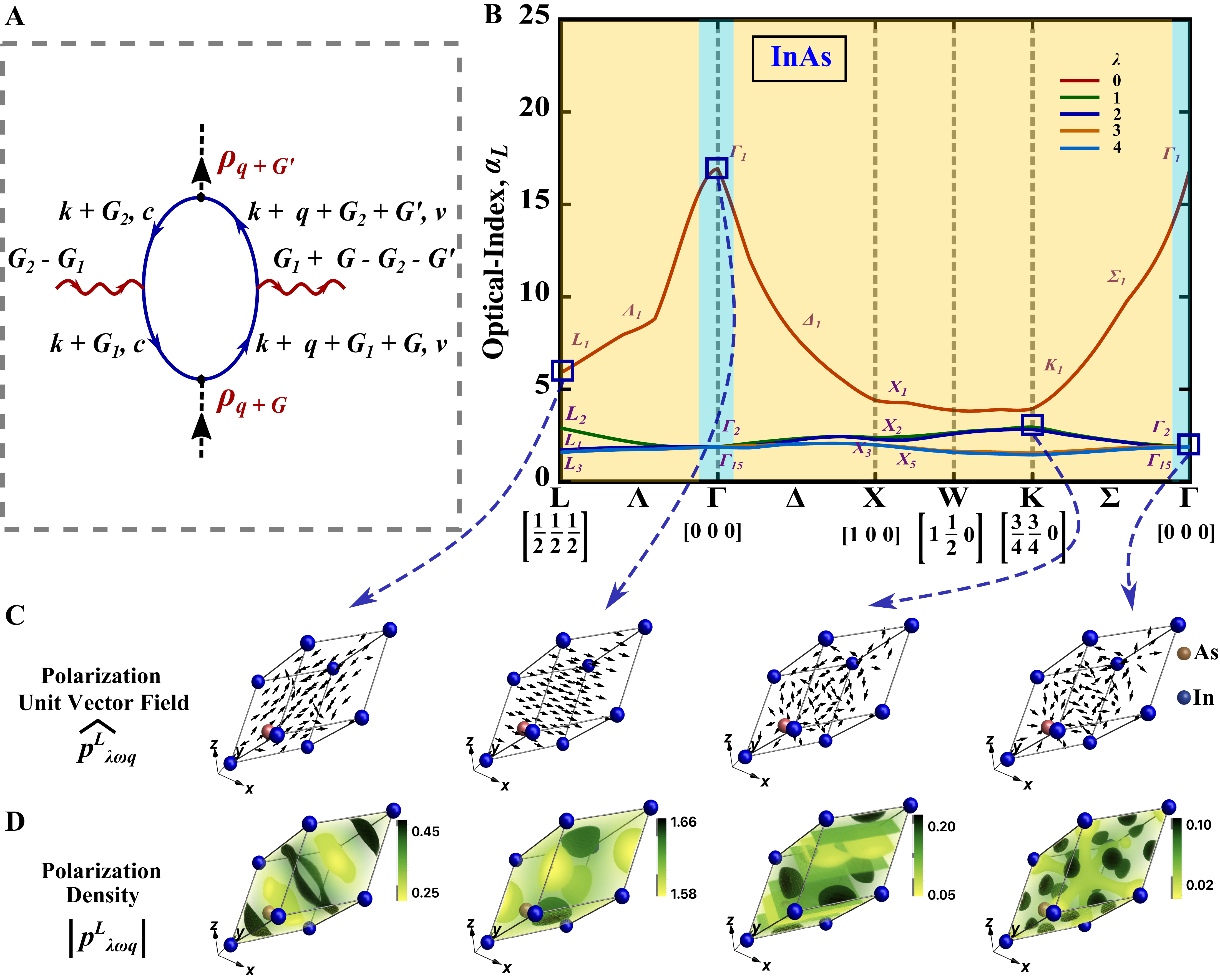}
    \caption{{\bf Unraveling hidden waves and deep microscopic optical waves (longitudinal) of a crystalline material}. We observe a significant texture and crowding of the hidden waves even within the primitive cell of a Zinc-Blend material. (A) The Feynman diagram depicts the lattice longitudinal conductivity of a crystalline material responsible for inducing the optical longitudinal polarization waves. The dashed lines represent the incoming ($\rho_{\bm{q}+\bm{G}}$) and outgoing ($\rho_{\bm{q}+\bm{G}'}$) charge density, internal solid lines are the electron propagators of the conduction ($c$) and valence ($v$) bands, and wiggly lines are the momentum exchange processes with the lattice. Here, $\bm{G},\bm{G}',\bm{G}_1,\bm{G}_2$ are the discrete reciprocal lattice vectors, and $\bm{q}$, $\bm{k}$ are restricted to the first Brillouin zone. (B) deep microscopic optical longitudinal band structure of InAs at $10\,$THz. The labels on the deep microscopic optical bands correspond to the irreducible representations of the $\bm{q}$-symmetry group. The blue-shaded region around the $\Gamma$-point shown here is the domain of classical optics. (C) Hidden optical longitudinal polarization unit vector field distribution at a few high-symmetry points plotted within the primitive cell of InAs. (D) Hidden optical longitudinal polarization density in units of $e^-/\AA^2$ corresponding to (C) plotted within a primitive cell of InAs.} 
    \label{fig:figure3}
\end{figure}

In Fig.~\ref{fig:figure3}(B), we plot the deep microscopic optical longitudinal hidden waves of InAs at a given frequency $\omega = 10\,$THz. The deep microscopic optical longitudinal waves determine the electronic screening properties of the material perturbed by an applied potential. In movies S3 and S4, we plot the first three optical hidden waves (transverse and longitudinal)  within a primitive lattice of InAs up to time $T = 0.05\,$ps, respectively. We observe that the hidden waves exhibit distinct temporal dynamics as they traverse through the crystal. 

We have studied topologically trivial optical materials in this study, which inherit only positive optical-indices. However, we note that the optical polarization waves hide unique topological indices such as the optical $N$-invariant not captured by electronic band theory. We note that the SILC is precisely the Green’s function of the polarization density of the material system. Following the Volovik formalism \cite{volovik2003universe, WangZhang}, we obtain the topological invariant corresponding to SILC that is conserved under continuous deformations. In two dimensions, for frequencies $\omega$ below the bandgap, we obtain the optical $N$-invariant from the deep microscopic optical band structure  (see supplementary information), given by 
\begin{equation}
N = \frac{1}{4\pi} \int d\bm{q} \sum_{\lambda = 0}^{D_f-1}{\rm sign}\left(\alpha_{\lambda\omega\bm{q}}\right)\bm{F}_{\lambda},
\end{equation}
where, $\bm{F}_{\lambda}$ is the Berry curvature of the optical polarization wave $\bm{p}_{\lambda\omega\bm{q}}$. From the above expression, we note that the crystals can exhibit a non-trivial optical $N$-invariant \mbox{$(N \neq 0)$} only when the optical indices can be continuously deformed from positive to negative. To find materials with non-trivial optical topological properties, it is therefore crucial to determine the deep microscopic optical band structure of a given material.  

\subsection*{Purdue-Picomax}
\begin{figure}
    \centering
    \includegraphics[width = 5.5in]{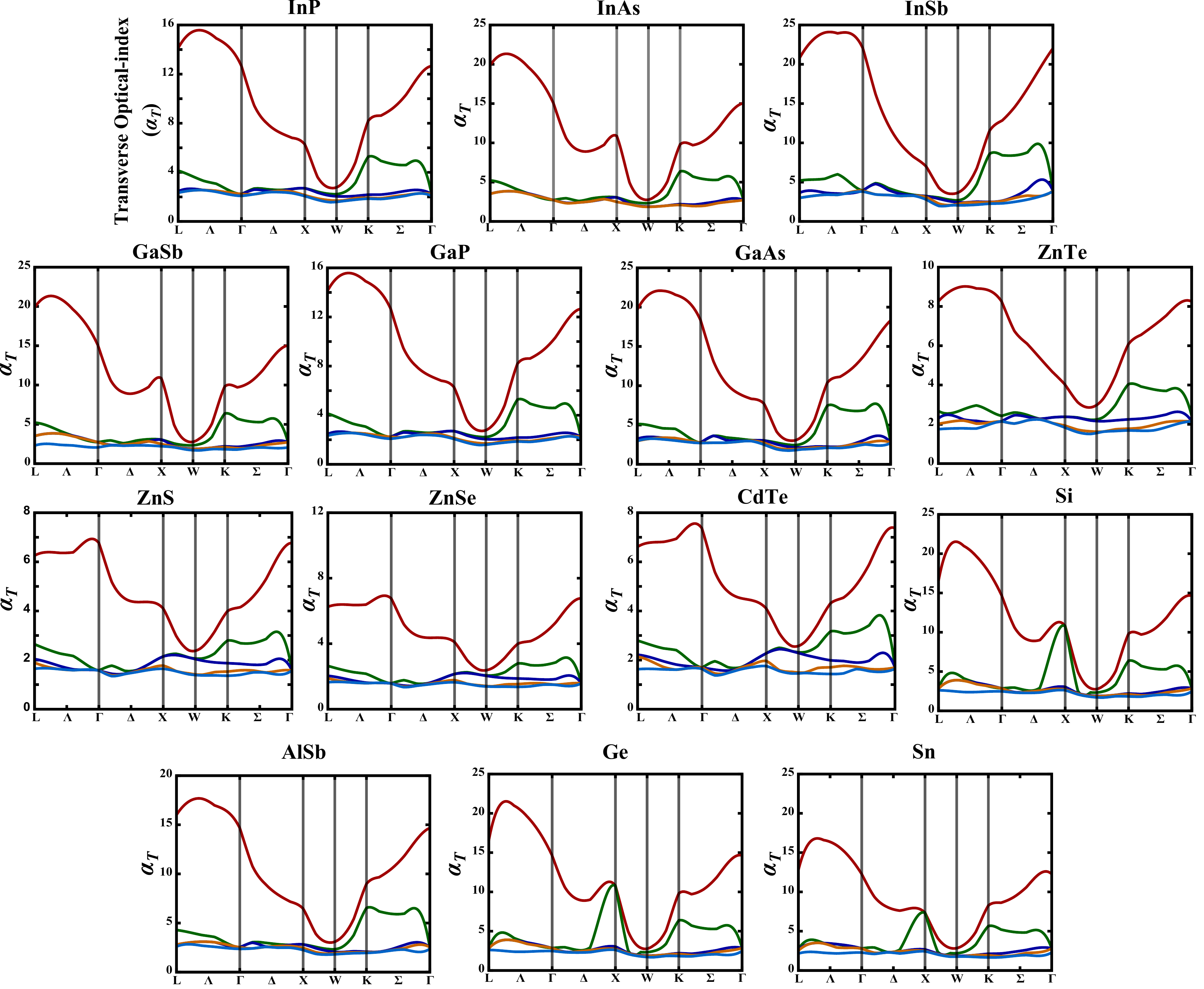}
    \caption{{\bf Deep microscopic hidden optical waves and optical transverse band structure of Zinc-Blend and Diamond Materials at 10 THz}. Unlike a single refractive index at 10 THz, our quantum theory provides a family of functions obeying crystal symmetries. These show the existence of hidden waves with specific momentum dispersion throughout the Brillouin zone. The curves (red, blue, green, etc.) correspond to different allowed bands of waves inside the crystalline material. We note that the deep microscopic optical transverse band structure of a material is crucial to determining the momentum direction with maximal optical response. For all the materials considered here, the first optical transverse index $\alpha^T_{0\,\omega\,\bm{q}}$ has a maximum at a momentum along the $\Lambda$ axis.}
    \label{fig:figure4}
\end{figure}

\begin{figure}
    \centering
    \includegraphics[width = 5.5in]{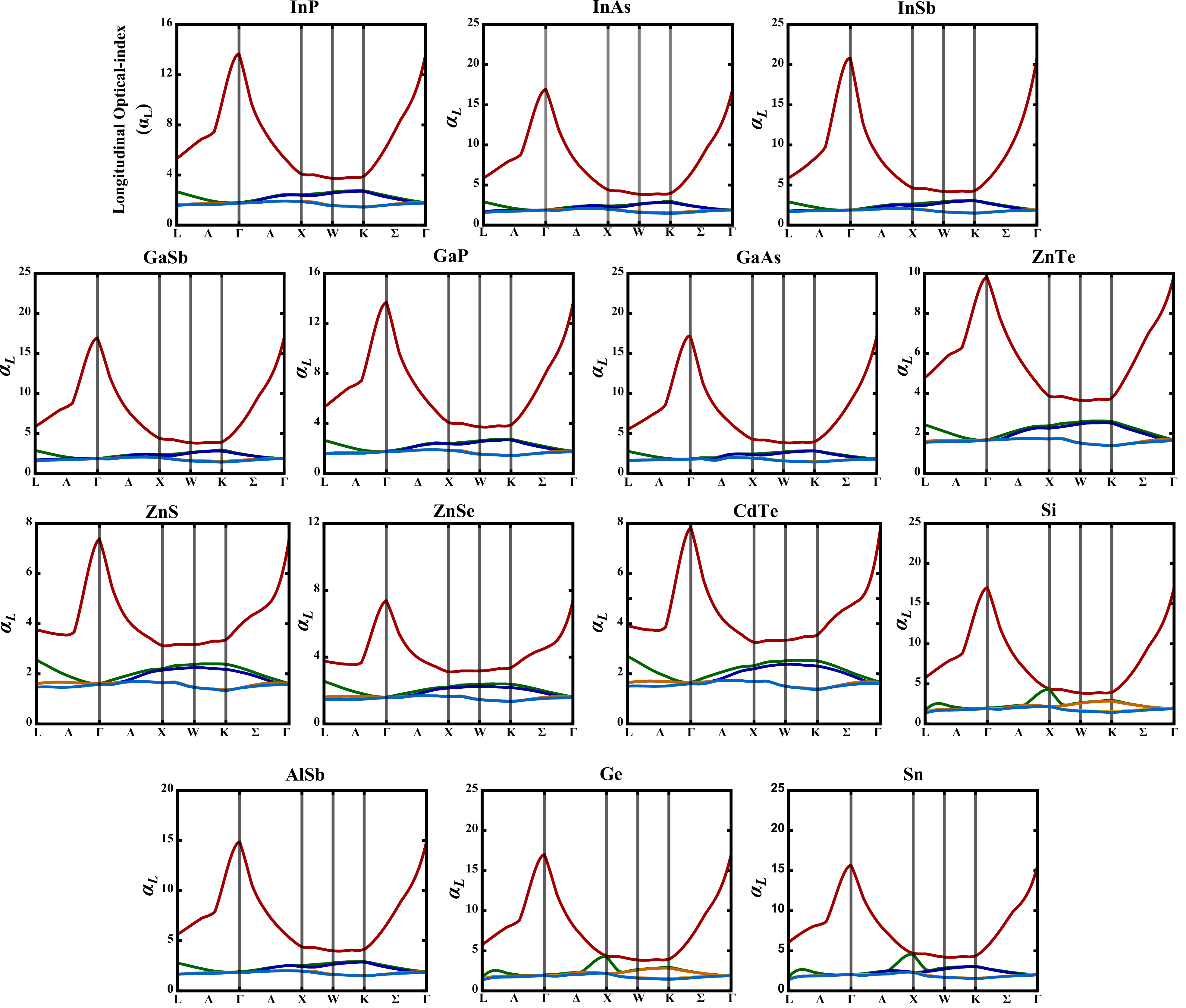}
    \caption{{\bf Deep microscopic hidden optical waves and optical longitudinal band structure of Zinc-Blend and Diamond Materials at 10 THz}. The deep microscopic optical longitudinal band structure determines the screening properties of a material. All possible bulk plasmon modes that a material can support are also determined by the optical longitudinal bands.}
    \label{fig:figure5}
\end{figure}

We have developed an open-source software, Purdue-Picomax, for the community to explore the deep microscopic optical band theory of solids. Purdue-Picomax computes unique physical quantities arising in our light-matter interaction theory. These include the symmetry-indicating lattice conductivity, deep microscopic optical transverse and longitudinal band structure, and hidden optical waves in various materials. The development of versatile electronic structure codes such as VASP \cite{VASP}, Quantum ESPRESSO \cite{QuantumESPRESSO} and others, and the phonon band structure codes such as EPW \cite{EPW}, ABINIT \cite{ABINIT}, Perturbo \cite{Perturbo} have revolutionized material science to design and discover novel materials with tunable electronic and phonon properties, respectively. We anticipate that the development of Purdue-Picomax facilitates a similar acceleration towards the discovery of materials with novel deep microscopic optical topological and electrodynamic phases of matter. Users from both the physics and material science communities can explore emerging materials for further research activities. Currently, we have included Group IV, III-V, and II–VI crystalline materials and the software can be easily adapted to include many more user-defined materials (eg: Moire materials). In Fig.~\ref{fig:figure4}, and Fig.~\ref{fig:figure5}, we have plotted the hidden waves (deep microscopic optical transverse and longitudinal polarization band structure) obtained using Purdue-Picomax for $14$ technologically important semiconducting materials. Purdue-Picomax is free software available on GitHub \mbox{\url{https://github.com/sathwik92/PicoMax1.0}} under MIT license.

\section{Discussion}
We have unraveled the deep microscopic optical band structure of crystalline materials. A complete description of the optical polarization waves in a crystal lattice is encoded in the deep microscopic optical band structure. We have shown that the optical polarization waves display a rich spatial and temporal distribution, hence representing a significant polarization texture and crowding even within a primitive lattice of a material. 

The constant development of new near-field tools for light-matter interaction makes it an exciting frontier within reach to probe these hidden waves. One can find signatures of the optical longitudinal waves from coherent resonant inelastic scattering \cite{schulke2007electron} or from the momentum-resolved electron energy loss experiments \cite{poursoti2022deep}  along different crystal orientations (see Supplementary Materials). The deep microscopic optical transverse band structure can be measured via momentum-resolved Fourier transform reflectometry \cite{decrescent2016model} along different crystal growth directions (see Supplementary Materials). The deep microscopic optical transverse band structure of a material determines the momentum direction with maximal optical response. On the other hand, the optical longitudinal polarization waves determine the Coulombic screening properties of the material. The optical longitudinal bands also encode the multi-band plasmon dispersion \cite{papaj2023probing, husain2023pines}, a key signature associated with the correlated phases in materials. Plasmon frequencies for a given crystal momentum $\bm{q}_0$ can be inferred by recognizing the maxima of the polarization loss function 
$L_{\lambda\,\bm{q}_0\,\omega} = {\rm Im}\left(-1/\alpha_{\lambda\,\bm{q}_0\,\omega}\right)$, corresponding to each optical longitudinal polarization band. 

As with the electron/phonon band theory of solids, we expect Purdue-Picomax and our deep microscopic optical band theory of solids to provide a framework to probe optical materials and optical topological phenomena with unique polarization textures at the lattice level.  Our findings point to a new crystallographic feature that serves as a generic framework for investigating the universal physics of lattice optical response of materials.   
\section{Methods}
We determined the electronic Bloch functions and the eigen-energies by the pseudo-potential method (see supplementary information). Symmetry indicating lattice conductivity components are determined by Eq.~\ref{eq:transversecond} \& \ref{eq:longitudinalcond}, respectively. We performed the eigenvalue decomposition in Eq.~\ref{eq:spectraldecomposition} to obtain the optical indices and the corresponding hidden waves. Shifted Monkhorst–Pack \mbox{$k$-points} mesh of $1120$ points were used in all our calculations for $\sigma_{\omega\,\bm{q}}^{\bm{G}\,\bm{G}'}$.

Our approach here is general and applicable to all optical frequencies and permissible momentum values throughout the Brillouin zone. As a limiting case at $\omega = 0$, the deep microscopic optical longitudinal band structure can also be employed in chemical compounds to identify the molecular screening properties. Classically, the longitudinal polarization at a given frequency is a constant field directed along $\hat{\bm{q}}$. We observe that the first optical longitudinal polarization wave around $\Gamma$-point (considered along $\Delta$ direction), $\bm{p}_{0\,\omega\,\,\bm{q}}^L$ is aligned along $\hat{\bm{q}} = [100]$ (see Fig.~\ref{fig:figure3}(C)), and the corresponding polarization density is nearly constant as shown in Fig.~\ref{fig:figure3}(D). Again a significant spatial texture of the optical longitudinal polarization is observed away from the $\Gamma$ point (Fig.~\ref{fig:figure3}(D), movie S2).  In Fig.~\ref{fig:figure3}(C), we observe that the first optical longitudinal polarization band $\alpha^L_{0\,\omega\,\bm{q}}$ has a maximum at the $\Gamma$-point and is significantly damped away from the $\Gamma$ point. This is attributed to the Coulombic screening which decays as $\sim\,1/|\bm{q}|^2$. 

We show that a key feature of the deep microscopic optical band theory of the crystalline materials is a distinct degeneracy pattern exhibited by the optical indices. The degeneracy pattern of optical polarization waves and the structure of SILC indicate symmetry throughout the Brillouin zone of the material. At the $\Gamma$ point, the optical polarization waves adopt the $T_{d}$ point group symmetry, and the first deep microscopic optical index at $\Gamma$-point $(\alpha^{L}_{0\,\omega\,\Gamma}, \alpha^{T}_{0\,\omega\,\Gamma})$ transforms as per the one-dimensional irreducible representation $\Gamma_1$. The next four optical indices belong to the irreducible representation $\Gamma_2$ and $\Gamma_{15}$ with degeneracy one and three, respectively. Going from the $\Gamma$ to $\rm L$-point, $T_d$ point group symmetry reduces to $C_{3V}$ symmetry. The optical polarization waves in the class of $\Gamma_{15}$ branch out into two deep microscopic optical bands with degeneracy two $(L_3)$ and one $(L_1)$ (see Fig.~\ref{fig:figure2}(B) and Fig.~\ref{fig:figure3}(B)). Similarly, the optical polarization waves throughout the Brillouin zone can be classified into a set of irreducible representations from the corresponding $\bm{q}$-symmetry group as shown in Fig.~\ref{fig:figure2}(B) and Fig.~\ref{fig:figure3}(B). 

\section*{Associate Content}
\subsection*{Supplementary Information}
Additional theoretical details on the symmetry indicated lattice conductivity and the implications of group symmetry in determining the lattice conductivity structure. Theoretical details on the pseudopotential method employed here for electronic band structure calculations. Suggested experimental strategies for detecting the deep microscopic optical band structure via momentum-resolved Fourier transform reflectometry and momentum-resolved electron energy loss spectroscopy. \\
Complex deep microscopic optical longitudinal and transverse band structure of InAs at frequency $\omega = 200, 300, 400, 500\,$THz (Figure S1).\\
Spatial evolution of the hidden optical transverse polarization wave (Movie S1).\\
Spatial evolution of the hidden optical longitudinal polarization wave (Movie S2).\\
Temporal evolution of the hidden optical transverse polarization waves (Movie S3).\\
Temporal evolution of the hidden optical longitudinal polarization waves (Movie S4).

\section*{Author contributions statement}
S.B. and Z.J. developed the idea. S.B. led and Z.J. supervised the project. All authors contributed to the production and analysis of the results. All authors reviewed the manuscript.

\section*{Competing interests}
The authors declare no competing interests.

\section*{Data and code availability}
The data and the code that support the findings of this study are available in the following GitHub repository
\mbox{\url{https://github.com/sathwik92/picomax1.0}} or from the corresponding author (\url{zjacob@purdue.edu}).

\begin{acknowledgement}
This work was supported by the Office of Naval Research (ONR) under the award no. 13001334.
\end{acknowledgement}

\bibliography{scibib}

\newpage
\section*{Supplementary Material: Unraveling Optical Polarization at Deep Microscopic Scales in Crystalline Materials}

\section*{Symmetry Indicated Lattice Conductivity (SILC)} 
Lattice conductivity in a crystal satisfies the translational invariance of the form
\begin{equation}
{\bm{\sigma}}(\bm{r}, \bm{r}', \omega) = {\bm{\sigma}}(\bm{r}+\bm{R}_n, \bm{r}'+\bm{R}_n, \omega),    
\end{equation} 
where, $\bm{R}_n$ is the translation vector in real space. Hence, we can expand ${\bm{\sigma}}(\bm{r}, \bm{r}', \omega)$ in the reciprocal lattice basis as
\begin{equation}\label{eq:latticecond}
{\bm{\sigma}}(\bm{r}, \bm{r}', \omega) = \frac{1}{\Omega}\sum_{\bm{q}, \bm{G}, \bm{G}'} \displaystyle e^{-i(\bm{q}+\bm{G}')\cdot\bm{r}'}\ \bm{\sigma}^{\bm{G}\,{\bm{G}}'}_{\omega\,\bm{q}} \ \displaystyle e^{i(\bm{q}+\bm{G})\cdot\bm{r}},
\end{equation}
where, $\Omega$ is the crystal volume. In a crystal, lattice conductivity also satisfies the invariance under space group operations ($\bm{\mathcal{G}}$), given by
\begin{equation}
{\bm{\sigma}}(\bm{\mathcal{G}}\cdot\bm{r}+\bm{\tau}_{\bm{\mathcal{G}}}, \bm{\mathcal{G}}\cdot\bm{r}'+\bm{\tau}_{\bm{\mathcal{G}}}, \omega) = {\bm{\sigma}}(\bm{r}, \bm{r}', \omega),
\end{equation}
where, the nonprimitive translation vector $\bm{\tau}_{\bm{\mathcal{G}}}$ is introduced to include the class of nonsymmorphic crystals. In the reciprocal lattice space, the space group operations imply the condition
\begin{equation}\label{eq:symmetryconductivity}
\bm{\sigma}^{\bm{G}\,{\bm{G}}'}_{\omega\,\bm{q}_t} = e^{i(\bm{G}'-\bm{G})\cdot\bm{\tau}_{\bm{\mathcal{G}}}}\,\bm{\sigma}^{\bm{G}_t\,{\bm{G}_t}'}_{\omega\,\bm{q}},
\end{equation}
where, $\bm{q}_t = \bm{\mathcal{G}}\cdot\bm{q} + \bm{G}_{\bm{\mathcal{G}}}$, $\bm{G}_t = \bm{\mathcal{G}}^{-1}\cdot(\bm{G}+\bm{G}_{\bm{\mathcal{G}}})$, and $\bm{G}_{\bm{\mathcal{G}}}$ is the auxiliary reciprocal lattice vector. The reality condition dictates that
\begin{equation}
\bm{\sigma}^{\bm{G}\,{\bm{G}}'}_{\omega\,\bm{q}} = {\bm{\sigma}^*}^{-\bm{G}\,{-\bm{G}}'}_{-\omega\,-\bm{q}}.
\end{equation}
Hence, the lattice conductivity of a crystalline material is symmetry indicating, and the structure of the lattice conductivity is determined by the space group symmetry for a given crystal momentum $\bm{q}$. This relation implies that the lattice optical response in solids belongs to the universality \mbox{class-$\mathcal{D}$} of real bosons. In topological superconductors, this is typically referred to as the particle-hole \mbox{$\mathcal{C}$-symmetry}. $\mathcal{C}$-symmetry is an exact symmetry in the lattice optical response, however only an approximate in superconductors.

The space-group symmetry dictated by $\bm{q}$ within the crystal lattice determines the structure of the lattice conductivity, and the corresponding symmetry transformation is given in Eq.~\ref{eq:symmetryconductivity}. In zinc-blend materials, at $\rm \Gamma$-point, the symmetry group is $T_d$. The point group $T_d$ has $24$ elements and the degree of degeneracy for the optical polarization modes can be up to three as per the irreducible representations. In our calculations, we have considered nine $\bm{G}$ vectors to result in a nine-dimensional tensor for the lattice conductivity. Let us consider, $\left\{\bm{G}_i\right\}_{i = 0}^{9} = \left\{ [000], [\Bar{1}11], [1\Bar{1}1], [11\Bar{1}], [111], [1\Bar{1}\Bar{1}], [\Bar{1}1\Bar{1}] , [\Bar{1}\Bar{1}1], [\Bar{1}\Bar{1}\Bar{1}]\right\}$ in units of $(2\pi/a)$. Through symmetry transformations, we obtain 9 independent components in the conductivity tensor and the corresponding expression is given by
\begin{equation}\label{eq:gammapointcond}
\bm{\sigma}_{\omega\,\Gamma}=\left(
\begin{array}{ccccccccc}
 \sigma_{00} & \sigma_{01} & \sigma_{01} & \sigma_{01} & \sigma_{04} & \sigma_{04}
   & \sigma_{04} & \sigma_{04} & \sigma_{01} \\
 \sigma_{01} & \sigma_{11} & \sigma_{12} & \sigma_{12} & \sigma_{14} & \sigma_{15}
   & \sigma_{14} & \sigma_{14} & \sigma_{12} \\
 \sigma_{01} & \sigma_{12} & \sigma_{11} & \sigma_{12} & \sigma_{14} & \sigma_{14}
   & \sigma_{15} & \sigma_{14} & \sigma_{12} \\
 \sigma_{01} & \sigma_{12} & \sigma_{12} & \sigma_{11} & \sigma_{14} & \sigma_{14}
   & \sigma_{14} & \sigma_{15} & \sigma_{12} \\
 \sigma_{04} & \sigma_{14} & \sigma_{14} & \sigma_{14} & \sigma_{44} & \sigma_{45}
   & \sigma_{45} & \sigma_{45} & \sigma_{15} \\
 \sigma_{04} & \sigma_{15} & \sigma_{14} & \sigma_{14} & \sigma_{45} & \sigma_{44}
   & \sigma_{45} & \sigma_{45} & \sigma_{14} \\
 \sigma_{04} & \sigma_{14} & \sigma_{15} & \sigma_{14} & \sigma_{45} & \sigma_{45}
   & \sigma_{44} & \sigma_{45} & \sigma_{14} \\
 \sigma_{04} & \sigma_{14} & \sigma_{14} & \sigma_{15} & \sigma_{45} & \sigma_{45}
   & \sigma_{45} & \sigma_{44} & \sigma_{14} \\
 \sigma_{01} & \sigma_{12} & \sigma_{12} & \sigma_{12} & \sigma_{15} & \sigma_{14}
   & \sigma_{14} & \sigma_{14} & \sigma_{11} \\
\end{array}
\right).
\end{equation}
At $\rm L$-point, the point group symmetry $C_{3V}$ results in 18 independent components in the conductivity tensor, given by
\begin{equation}\label{eq:lpointcond}
\bm{\sigma}_{\omega\,L}=\left(
\begin{array}{ccccccccc}
 \sigma_{00} & \sigma_{01} & \sigma_{01} & \sigma_{01} & \sigma_{04} & \sigma_{05}
   & \sigma_{05} & \sigma_{05} & \sigma_{08} \\
 \sigma_{01} & \sigma_{11} & \sigma_{12} & \sigma_{12} & \sigma_{14} & \sigma_{15}
   & \sigma_{16} & \sigma_{16} & \sigma_{18} \\
 \sigma_{01} & \sigma_{12} & \sigma_{11} & \sigma_{12} & \sigma_{14} & \sigma_{16}
   & \sigma_{15} & \sigma_{16} & \sigma_{18} \\
 \sigma_{01} & \sigma_{12} & \sigma_{12} & \sigma_{11} & \sigma_{14} & \sigma_{16}
   & \sigma_{16} & \sigma_{15} & \sigma_{18} \\
 \sigma_{04} & \sigma_{14} & \sigma_{14} & \sigma_{14} & \sigma_{44} & \sigma_{45}
   & \sigma_{45} & \sigma_{45} & \sigma_{48} \\
 \sigma_{05} & \sigma_{15} & \sigma_{16} & \sigma_{16} & \sigma_{45} & \sigma_{55}
   & \sigma_{56} & \sigma_{56} & \sigma_{58} \\
 \sigma_{05} & \sigma_{16} & \sigma_{15} & \sigma_{16} & \sigma_{45} & \sigma_{56}
   & \sigma_{55} & \sigma_{56} & \sigma_{58} \\
 \sigma_{05} & \sigma_{16} & \sigma_{16} & \sigma_{15} & \sigma_{45} & \sigma_{56}
   & \sigma_{56} & \sigma_{55} & \sigma_{58} \\
 \sigma_{08} & \sigma_{18} & \sigma_{18} & \sigma_{18} & \sigma_{48} & \sigma_{58}
   & \sigma_{58} & \sigma_{58} & \sigma_{88} \\
\end{array}
\right).
\end{equation}
A similar analysis is performed for the $\bm{q}$-symmetry groups over the entire Brillouin zone to obtain the optical lattice conductivity tensor and the corresponding polarization waves. Diamond crystal lattice belongs to the nonsymmorphic space group, and the symmetry considerations are distinct from the zinc-blend materials. Especially at $\rm X$-point, symmetry considerations allow only double-degenerate hidden optical polarization waves (as shown in Fig.~4 and Fig.~5 of the main text) for Si, Ge, and Sn. 
\subsection*{Empirical Pseudopotential Method for Electronic band structure}
Energy eigenvalues $(\epsilon_{nk})$ and the wavefunctions $\left|n,\bm{k}\right>$ corresponding to the electronic band structure are required to calculate $\sigma_{\omega\,\bm{q}}^{\bm{G}\,\bm{G}'}$. We obtain the electronic band structure based on the empirical pseudopotential method. Hamiltonian of a solid within this framework is given by
\begin{equation}
H_{\bm{K}\bm{K}'} = \frac{\hbar^2 \left|\bm{k}+\bm{K}\right|^2}{2m} \delta_{\bm{K}\bm{K}'} + V_L(|\bm{K}-\bm{K}'|) + V_{NL}(\bm{K}, \bm{K}'),
\end{equation}
where, $\bm{K}, \bm{K}'$ are reciprocal lattice vectors, $\bm{k}$ is the electronic wavevector, $V_L$ is the local pseudopotential and $V_{NL}$ is the nonlocal pseudopotential. 
\subsection*{Experimental Detection of deep microscopic optical Band Structure}
In this section, we discuss the experimental considerations for the observation of the deep microscopic optical band structure discussed in this work. The deep microscopic optical longitudinal band structure can be inferred from the coherent resonant inelastic scattering \cite{poursoti2022deep} or momentum-resolved electron energy loss experiments \cite{decrescent2016model} along different crystal orientations by determining the double differential cross-section. Since the inelastic scattering experiments involve a two-photon process, from the Kramers–Heisenberg formula, we see that contributions from the $|\bm{A}|^2$ term in the interaction Hamiltonian has the most dominant contribution. Hence, provides a pathway to measure the deep microscopic optical longitudinal band structure. The key quantity to obtain in these measurements is the lattice nonlocal structure factor, defined as
\begin{equation}
S_{\bm{q}}(\bm{G}, \bm{G}', \omega) = -\frac{|\bm{q}+\bm{G}||\bm{q}+\bm{G}'|}{4\pi e^2}\,{\rm Im}\left[\frac{1}{\delta_{\bm{G}\bm{G}'}+\left(\frac{4\pi i(\sigma^L)^{\bm{G}\bm{G}'}_{\omega\bm{q}}}{\omega}\right)}\right]. 
\end{equation}
In a coherent resonant inelastic scattering experiment, an incoming field with wavevector $\bm{k}_{in}$ is incident on the sample and both the Bragg reflected field with wavevector $\bm{k}_{\rm bragg}$, and scattered field with wavevector $\bm{k}_{\rm sc}$ can be measured. From the Bragg condition, we obtain \mbox{$\bm{G}' = \bm{k}_{\rm bragg}-\bm{k}_{\rm in}$}, and from the scattered field we can obtain $\bm{q}+\bm{G} = \bm{k}_{\rm in} - \bm{k}_{\rm sc}$. The lattice nonlocal structure factor $S_{\bm{q}}(\bm{G}, \bm{G}', \omega)$ can be obtained with the measurement of double differential cross-section. Hence, one can experimentally obtain the full lattice conductivity at a given frequency, and further resolve the deep microscopic optical longitudinal band structure.

The deep microscopic optical transverse band structure can be measured via momentum-resolved Fourier transform reflectometry \cite{decrescent2016model} along different crystal growth directions. The angle-dependent study provides a tool to tune the wavevector $\bm{q}$ along high-symmetry directions of the Brillouin zone. The incident field generates the current-current correlation and the measured reflectance spectra can be used to obtain the optical-index $\alpha_{\lambda\,\omega\,\bm{q}}$. A detailed analysis of the consequences of such measurements will be considered in future work.
\begin{figure}
    \centering
    \includegraphics[width = 5.5in]{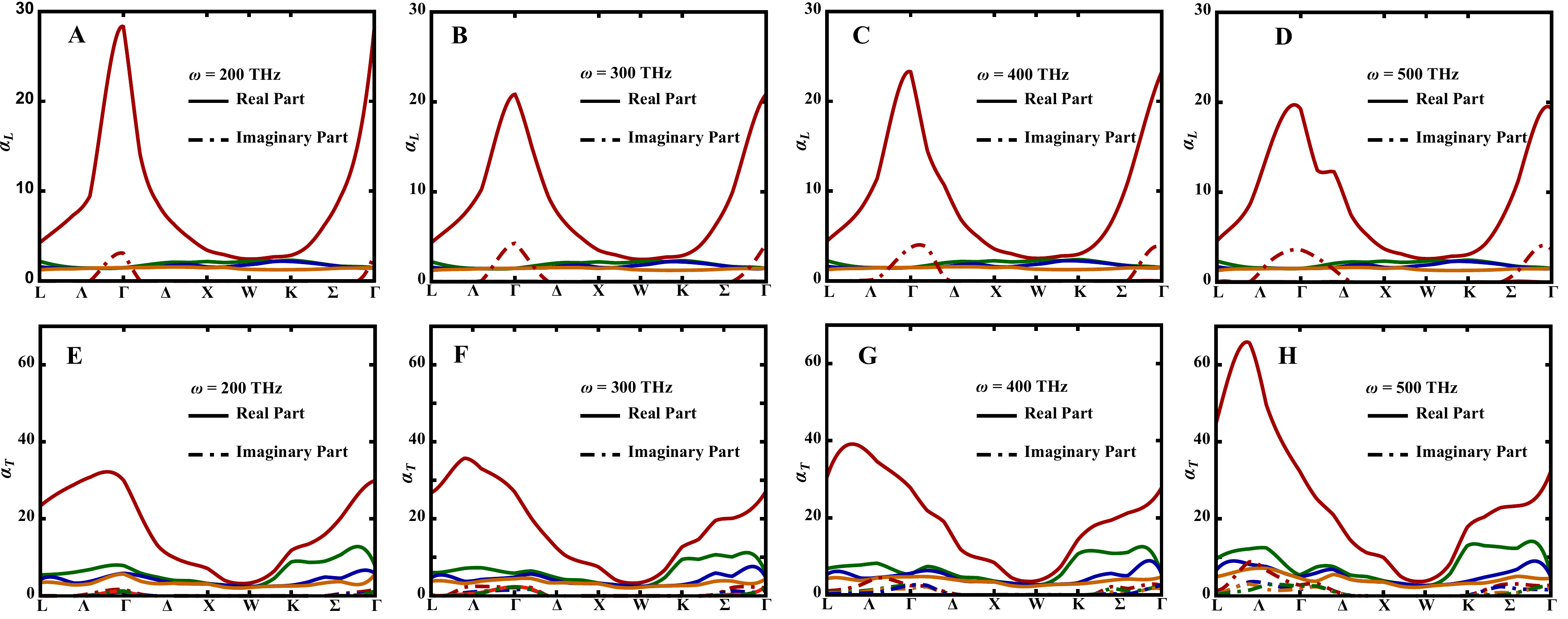}
    \caption{{\bf Complex deep microscopic optical longitudinal and transverse band structure}. Deep microscopic optical longitudinal (A, B, C, D) and transverse (E, F, G, H) band structure in InAs at frequency $\omega = 200, 300, 400, 500\,$THz.} 
    \label{fig:suppfig3}
\end{figure}



\section*{Topological Invariant for deep microscopic optical Band Theory of Solids}
The symmetry indicating lattice conductivity can be thought of as a Green's function of the system. Hence, we are interested in obtaining the topological invariant corresponding to SILC. For simplicity here we consider a 2+1D system. Following a familiar recipe from the topological band theory, the optical topological invariant corresponding to a continuous deformation of SILC is given by:
\begin{equation}
N = \frac{\epsilon^{ijk}}{24\pi^2} \int \int d\Omega\, d\bm{q}\, {\rm tr}\left(\sigma_{\bm{q}\omega}^{\bm{G}\bm{G}'}\,\frac{\partial\left(\sigma_{\bm{q}\omega}^{\bm{G}\bm{G}'}\right)^{-1}}{\partial q_i}\sigma_{\bm{q}\omega}^{\bm{G}\bm{G}'}\,\frac{\partial\left(\sigma_{\bm{q}\omega}^{\bm{G}\bm{G}'}\right)^{-1}}{\partial q_j}\sigma_{\bm{q}\omega}^{\bm{G}\bm{G}'}\,\frac{\partial\left(\sigma_{\bm{q}\omega}^{\bm{G}\bm{G}'}\right)^{-1}}{\partial q_k}\right), 
\end{equation}
where, $\Omega$ is the frequency analytically continued to the complex plane. Following the procedure of Wang and Zhang \cite{WangZhang}, we can circumvent the integral over the complex frequency space by expressing the above topological invariant in terms of the deep microscopic optical band structure parameters. We obtain the expression:
\begin{equation}
N = \frac{1}{4\pi} \int d\bm{q} \sum_{\lambda}{\rm sign}\left(\alpha_{\lambda\omega\bm{q}}\right)\bm{F}_{\lambda},
\end{equation}
where $\bm{F}_{\lambda}$ is the Berry curvature of the optical polarization wave $\bm{p}_{\lambda\omega\bm{q}}$, given by
\begin{equation}
\bm{F}_\lambda = -i\epsilon^{ij} \sum_{\bm{G}} \int \frac{dq_z}{2\pi}\, \partial_{q_i}\bm{p}^\dagger_{\lambda\omega\bm{q}}\cdot \partial_{q_j}\bm{p}_{\lambda\omega\bm{q}}.
\end{equation}
The total Berry curvature and hence the optical topological invariant $N \neq 0$ only when the optical-indices can be continuously deformed from positive to negative. 
\\ \\
\noindent{Movie S1.} {\bf Spatial evolution of the hidden optical transverse polarization wave}. The optical transverse polarization wave density and unit vector field distribution for the first band in InAs obtained at $10\,$THz plotted along high symmetry lines. 

\noindent{Movie S2.} {\bf Spatial evolution of the hidden optical longitudinal polarization wave}. The optical longitudinal polarization wave density and unit vector field distribution for the first band in InAs obtained at $10\,$THz plotted along high symmetry lines. 

\noindent {Movie S3.} {\bf Temporal evolution of the hidden optical transverse polarization waves}. Optical transverse polarization waves at $\rm X$-point in InAs obtained at frequency $10\,$THz for first (Movie 2A), second (Movie 2B), and third (Movie 2C) bands. 

\noindent {Movie S4.} {\bf Temporal evolution of the hidden optical longitudinal polarization waves}. Optical longitudinal polarization waves around $\rm \Gamma$-point in InAs obtained at frequency $10\,$THz for first (Movie 1A), second (Movie 1B), and third (Movie 1C) bands. 

\end{document}